\documentclass[10pt]{iopart}

\usepackage{iopams}
\usepackage{graphics}
\usepackage{supertabular}
\usepackage{array} 
\usepackage{float}

\newcommand{\sfrac}[2]{\textstyle\frac{#1}{#2}}

\begin{document}

\title{Structure of {\rm $^{23}$Al} from a 
       multi-channel algebraic scattering model based on mirror symmetry}

  \author{P. R. Fraser$^{1,*}$, A. S. Kadyrov$^{1}$, K. Massen-Hane$^{1}$,
          K.~Amos$^{2,3}$, L.~Canton$^{4}$, S.~Karataglidis$^{2,3}$,
          D.~van~der~Knijff$^{2}$, and I.~Bray$^{1}$}
  \ead{* paul.fraser@curtin.edu.au}

\address{$^{1}$ Department of Physics, Astronomy and Medical Radiation
Sciences, Curtin University, GPO Box U1987, Perth 6845, Australia}
\address{$^{2}$ School  of Physics,  University of  Melbourne,
Victoria 3010, Australia}
\address{$^{3}$ Department of Physics, University of
Johannesburg, P.O. Box 524 Auckland Park, 2006, South Africa}
\address{$^{4}$ Istituto  Nazionale  di  Fisica  Nucleare,
Sezione di Padova, Padova I-35131, Italia}

\date{\today}

\begin{abstract}
  The proton-rich nucleus $^{23}$Al has a ground state just 123 keV
  below the proton drip-line, and as a result comparatively little is
  known experimentally about its properties, as with many such
  nuclei. Theoretical investigations have tended to model exclusively
  the ground and first one to three excited states known.  In this
  paper, we theoretically model most of the known spectrum,
  and predict what states may as yet be unobserved.  We use the
  multichannel algebraic scattering (MCAS) method to describe states
  as resonances of a valence proton coupled to a $^{22}$Mg rotor
  core. Six states with low-excitation energies and defined $J^\pi$
  are matched, and we make the first prediction of the properties of
  four others and propound the possible existence of several more.
\end{abstract}


\submitto{\JPG}

\ioptwocol


\section{Introduction}

The proton-rich nucleus $^{23}$Al was first discovered in
1969~\cite{Ce69}, but having only one bound state, 123~keV below the
one-proton emission threshold, comparatively little is known about its
properties. As a result, much experimental effort has been spent on it
in recent years~\cite{Wi88,Ca01,He07,Ga08,Al10,Sa11,Ki11,Ba11,Ba12},
filling in details of the low-energy resonant spectrum.  Several
positive-parity states have been measured, some of which with details
tentatively known. No negative parity states have been observed,
though several should exist at low energies.  Theoretical
investigation goes back many years, with an early example being
Sherr's examination of the Coulomb displacement energy of the mirror
pair $^{23}$Ne and $^{23}$Al. These he compared with those predicted
by a shell model, as a part of a broad survey~\cite{Sh77}. A more recent
shell model investigation was undertaken using the OXBASH shell-model
code~\cite{Ox86} and the USD family of
interactions~\cite{Yu14}. Theoretical treatment of this pair as
nucleon and mass-22 systems has been undertaken in the context of
asymptotic normalisation coefficients (ANCs) in capture
reactions~\cite{Ti05a,Ti11} (and references therein).

Most of those theoretical studies concentrated upon the Gamow window
region and so tend to focus only upon the ground and one or few
low excitation resonance states. There are more resonance states in
the known spectrum and we seek a description of this currently known
spectrum.

We use the multi-channel algebraic scattering (MCAS) method to predict
spectral properties of ${}^{23}$Al to $\sim$4~MeV excitation, and to
obtain elastic scattering cross sections. The method solves the
coupled-channel problem of a nucleon coupled to a number of states in
a `core' nucleus.  Lippmann-Schwinger equations for the
coupled-channel problem are solved in momentum space starting with a
set of coupled-channel interactions defined in coordinate space.  They
are specified using a collective model prescription. Those
coupled-channel interactions are expanded using a set of Sturmian
functions, and the coupled-channel Green's function is determined via
an iterative process~\cite{Am03}.

The coupled-channels scattering potentials of the system are formed
with a target described by a Tamura collective model of rotor
character~\cite{Ta65}, coupled to the projectile nucleus. The basic
potential used for is
\begin{eqnarray}
V_{c'c}(r) = f(r) \biggl\{ V_0 \delta_{c'c} + V_{ll} [ {\bf {\ell \cdot 
\ell}} ]_{c'c}
+ V_{ss} [ {\bf {s \cdot I}} ]_{c'c} \biggr\}\nonumber\\
\hspace{1.4cm} + g(r) V_{ls}
[{\bf {\ell \cdot s}}]_{c'c}\ ,
\label{www}
\end{eqnarray}
where $c$ and $c'$ denote channels, $V_{0}$ is the central potential
strength, $V_{\ell\ell}$ is the orbital angular momentum dependent
potential strength, $V_{\ell s}$ is the spin-orbit potential strength,
and $V_{ss}$ is the spin-spin potential strength, with all parameters
in MeV. The channels ($c$) denote the underlying set of unique quantum
numbers for each value of the total spin-parity of the compound system
($J^\pi$). The potentials are taken to have a Woods-Saxon form,
\begin{equation}
f(r) = \left[1 + \exp{\left( {{r-R_0}\over a_0} \right)} \right]^{-1}
\hspace*{0.3cm} ; \hspace*{0.3cm} g(r) = \frac{1}{r} \frac{df(r)}{dr}
\label{ws} 
\end{equation}
and matrix elements are shown in detail in Ref.~\cite{Am03}. This
potential is then expanded to second order in terms of deformation of
the nucleus. Full details of the expansion and the final form are
provided in the Appendix of Ref.~\cite{Fr14}.

To account for the effects of Pauli blocked orbitals, orthogonalizing
pseudo-potentials (OPP) are utilised. The OPP method was developed
from the Orthogonality Condition Model of Saito~\cite{Sa69}. The
original potential, $V_{c \, c'} (\boldsymbol{r})$, is modified by
adding OPP to define 
\begin{equation}
\mathcal{V}_{cc'} (\boldsymbol{r}, \boldsymbol{r}') = V_{cc'} (\boldsymbol{r}) \,
\delta(r - r') + \mathcal{\lambda}_{c} \, A_{c}(\boldsymbol{r}) 
A_{c'}( \boldsymbol{r}')\delta_{cc'} \, ,
\label{OPP}
\end{equation}
where $A_{c}( \boldsymbol{r})$ is the single-nucleon radial wave
function of an occupied orbital. For OPP to block the Pauli-forbidden
orbitals, the blocking parameter, $\lambda_{c}$ is set ostensibly to
infinity, but 10$^6$~MeV suffices in practice.  Though not used here,
OPP blocking energies can be set at values between zero and infinity,
for cases of `hindrance' from partially filled
orbitals~\cite{Ca06a}. Full details are presented in Ref.~\cite{Am13}.

Consequently, the MCAS approach is particularly suitable for studying
properties of the first few MeV of the spectra of systems such as
${}^{23}$Al, which may be considered in that regime as a nucleon plus
(core) nucleus clusters. Further, as the method readily gives
scattering amplitudes, it enables prediction of cross sections for
low-energy nucleon scattering from those `core' nuclei in which
resonance states of the compound can have effect.

To calculate a spectrum for ${}^{23}$Al, we first study the mirror
system, ${}^{23}$Ne, as a $n$+${}^{22}$Ne cluster so defining the
nuclear part of the $p$+${}^{22}$Mg Hamiltonian. Results for the
spectrum of ${}^{23}$Ne are given in Section~\ref{Ne} for
completeness. They differ slightly from those published
recently~\cite{Fr15}, as in that publication the model parameters were
tuned to the case where coupling to one less target state was
considered.

The nuclear interaction so defined is then used for the
$p$+${}^{22}$Mg system. In Section~\ref{charge} a discussion is
presented regarding the form of the Coulomb interactions added (with
an alternative form discussed in \ref{App-3pF}), and in
Section~\ref{Al} we give a prediction of the spectrum of
${}^{23}$Al. Cross sections for elastic scattering of protons from
${}^{22}$Mg from $E_{lab} =$ 0 to 3~MeV are also shown and discussed
therein. (Data at energies higher than the potential is designed for
are examined in \ref{App-Al2}.)  In Section~\ref{no2_2}, we present
results obtained when the number of target nuclear states, used in the
coupled-channel calculations, is reduced to that previously
deemed~\cite{Ba11} to contribute to the ground state of 23 Al.
Finally, in Section~\ref{Conclusion} we draw our conclusions.

\section{The nuclear interaction from the ${\rm n}$+${\rm ^{22}Ne}$ system}
\label{Ne}

In a recent paper~\cite{Fr14}, a good match was obtained between
experiment and MCAS calculation for the excitation energies of the
nine lowest eigenstates of $^{23}$Ne, treated as the $n+^{22}$Ne
system. These are all deeply bound with regard to the neutron emission
threshold. Those states span an energy range of $\sim$3.5~MeV. At
energies higher than this, processes other than coupling of a single
particle to a collective target with two rotor-like bands come into
effect.

The coupled channels for the mirror $n$+${}^{22}$Ne study were formed
using a rotational model description of the lowest five states of
${}^{22}$Ne. The three lowest energy states were taken to be states of
the principle rotor band; the $0^+_{g.s.}$, the $2^+_1$ state
(1.274~MeV in $^{22}$Ne and 1.247~MeV in $^{22}$Mg), and the $4^+_1$
state (3.357~MeV in $^{22}$Ne and 3.308~MeV in $^{22}$Mg). Next we add
the $4^+_2$ state at 5.523~MeV excitation in ${}^{22}$Ne (5.293~MeV in
${}^{22}$Mg). We include this state since, recently, the one-proton
knock-out reaction from the ground state of $^{23}$Al was studied
experimentally by Banu \textit{et al.}~\cite{Ba11}, with $\gamma$-rays
from the resultant excited $^{22}$Mg indicating which states of that
nucleus combine with a proton to populate the $^{23}$Al ground
state. Three $\gamma$-transitions were observed, $4^+_2 \rightarrow
4^+_1$, $4^+_1 \rightarrow 2^+_1$, and $2^+_1 \rightarrow 0^+_1$.
However, while this experiment suggests that only these four target
states are relevant to defining the $^{23}$Al ground state, we seek to
define more states than the ground state. Thus, the fourth known state
of the spectrum, the $2^+_2$ state (4.456~MeV in $^{22}$Ne and
4.402~MeV in $^{22}$Mg) was included. This state, like the $2^+_1$
state, $E2$-decays to the ground~\cite{Ba15}. In ${}^{22}$Mg, the
$4^+_2$ also $\gamma$-decays to the $0^+_{g.s.}$ ~\cite{Ba15}, $2^+_1$
(characterised as $M1$~\cite{Ba15} + $E2$~\cite{Fi05}) and
$4^+_1$~\cite{Ba15} states. Thus, we expect it to
play a role in channel coupling, even if not significantly with
regard to the ground state of $^{23}$Al. This is discussed in
Section~\ref{no2_2}.

\begin{figure*}[htp]
\begin{center}
\scalebox{0.55}{\includegraphics*{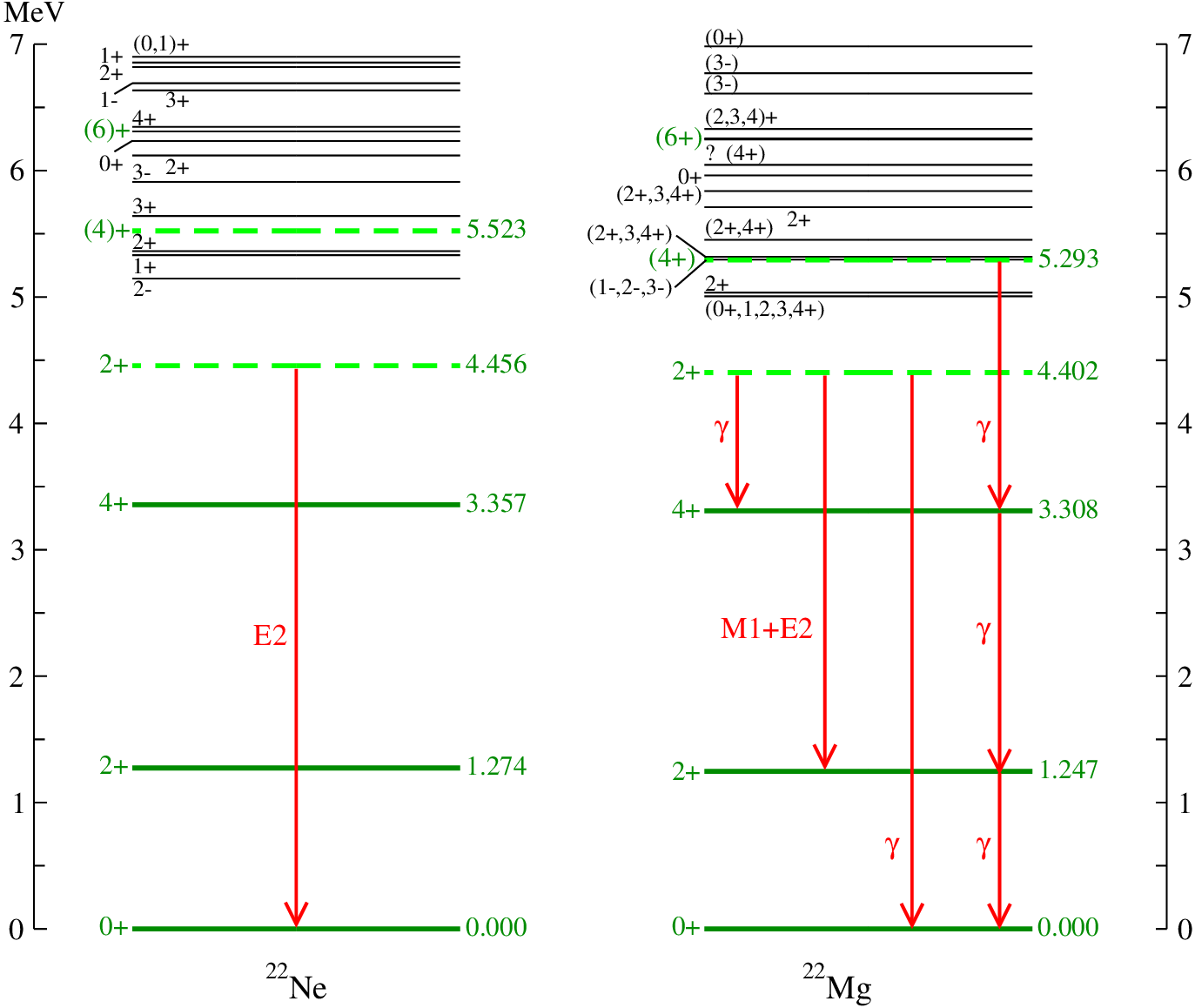}}
\end{center}
\caption{ \label{Ne22-spec} The low-energy experimental spectra of
  $^{22}$Ne and $^{22}$Mg. Solid, thick lines represent states of the
  main rotor band, and thick, dashed lines are states known to couple
  to them by $\gamma$-emission. Observed couplings between states in
  and out of the main rotor band shown by arrows. Data are from
  Ref.~\cite{Ba11,Fi05,Ba15}.}
\end{figure*}

All of the above experimentally-known $\gamma$-transitions are
summarised in Fig.~\ref{Ne22-spec}. We note that in the case of
$^{22}$Mg, most possible decays between these five states have
been observed experimentally, with four exceptions: there
have been no observed $4^+_2 \rightarrow 2^+_2$, $4^+_2 \rightarrow
2^+_1$, $4^+_2 \rightarrow 0^+_1$, or $4^+_1 \rightarrow 0^+_1$
transitions. In our calculations we still assume all couplings,
however.

The $2^+_2$ and $4^+_2$ states are not members of the main rotor band,
but they can still be considered using the rotational model. Assuming
a shape co-existence, these appear part of the second $K=0$
$\beta$-band, though no $0^+_2$ band head has been clearly
observed. Alternatively, they may be the first and third states of a
$K=2$ $\gamma$-band, in which case all states should possess the same
deformation, though no $3^+_2$ state, the second state in such a band,
has been clearly observed~\cite{Da68}. There is an
uncertainly-assigned state in $^{22}$Mg, denoted $(0^+,1,2,3,4^+)$, at
5~MeV that could be either~\cite{Fi05}.  Here, we effectively treat
these states as part of a secondary $K=0$ rotor band.

We denote the quadrupole coupling deformation of the primary and
secondary rotor band as $\beta_2$ and $\overline{\beta_2}$,
respectively. Assuming shape co-existence, we determine the ratio of
these two deformations by considering the mean lifetime of a state
that $\gamma$-decays to the ground state, which is a function of the
transition probabilities, and for $E2$ multipolarity this is:
\begin{equation}
\frac{1}{\tau}
= 1.23 \times 10^9\; E_\gamma^{\,5} \; B(E2) \,,
\label{taube2}
\end{equation}
with $\tau$ in s, $E_\gamma$ in MeV, and $B(E2)$ in $e^2$ fm$^4$. In
${}^{22}$Ne, the $E2$ transitions of the $2^+$ states at 1.275 and
4.456~MeV have half lives of 3.63(5)~ps and 37(6)~fs
respectively~\cite{Fi05}. Thus, from Eq.~(\ref{taube2}) we obtain
$B(E2)\bigr|_{1.275} = 46.06$ and $B(E2)\bigr|_{4.456} = 8.67$ $e^2$
fm$^4$.  The ratio of these values is 0.188, and given that, without
considering band quantum numbers and going only to the first order, as
$B(E2)$ values of a collective model with rotational character are
proportional to $\beta_2^2$, $\overline{\beta_2}~=~0.43\,\beta_2$.  As
the simplest possible formalism, we assume that couplings involving
the $4^+_2$ have the same $\overline{\beta_2}$.

\begin{figure}[htp]
\begin{center}
\scalebox{0.6}{\includegraphics*{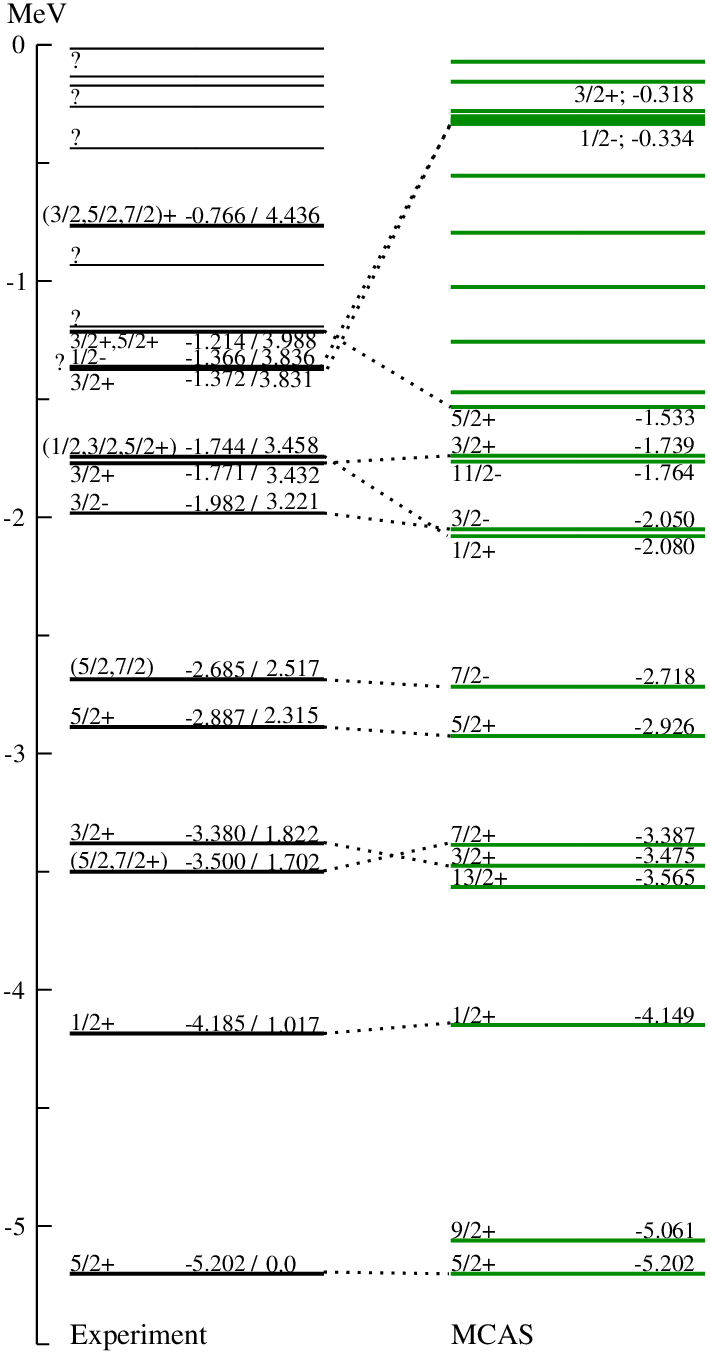}}
\end{center}
\caption{ \label{Ne23-spec} The experimental $^{23}$Ne
  spectrum~\cite{Fi07} and that calculated from MCAS evaluation of the
  $n+^{22}$Ne, with target state set $0^+_1$, $2^+_1$, $4^+_1$,
  $\overline{2^+_2}$ and $\overline{(4)^+_2}$.  The bar denotes the
  use of reduced coupling for channels involving this state. In the
  experimental spectrum, energies before the slash are relative to the
  one-proton emission threshold, and energies after are relative to
  the ground state.}
\end{figure}

Table~\ref{params} contains the parameters that define the nuclear
potential used in calculation of both $n+^{22}$Ne and $p+^{22}$Mg. The
positive-parity central well depth, $V_0$, differs from the value of
-51.3~MeV used in Ref.~\cite{Fr14}, to bring the $^{23}$Ne ground
state energy, relative to the neutron emission threshold, in line with
experiment to three decimal points. (In Ref.~\cite{Fr14}, the
parameters were selected without consideration of the $4^+_2$ state.
The adjustment here is to address that.) $\beta_4$ indicates a small
hexadecapole deformation, and $R_0$ and $a_0$ are radius and
diffusivity of Eq.~(\ref{ws}). The MCAS spectrum of $^{23}$Ne found
using the parameter set of Table~\ref{params} is compared to
experiment in Fig.~\ref{Ne23-spec}. This calculation results in a
match, within 0.5~MeV, for the nine lowest-energy states known for
which spin-parities are well assigned or postulated.  This makes a
good result for a range of up to 3.5~MeV.

\begin{table}\centering
\caption{\label{params} 
Parameter values defining the $n+^{22}$Ne interaction. $\lambda^{(OPP)}$
are blocking strengths of occupied shells, in MeV, with the same values used
for all target states considered.}
\begin{supertabular}{>{\raggedright}p{32mm} p{19mm}<{\centering} 
p{19mm}<{\centering}}
\hline
\hline
 & Odd parity & Even parity\\
\hline
$V_0$ (MeV) Fig.~\ref{Ne23-spec} \& \ref{Al23-spec-Wa07} & -65.200 & -50.894\\
$V_0$ (MeV) Fig.~\ref{Al23-spec-Wa07mod}                 & -64.620 & -50.372\\
$V_0$ (MeV) Fig.~\ref{Al23-spec-Wa07no2_2}               & -64.825 & -50.650\\
$V_0$ (MeV) Fig.~\ref{Al23-spec-He07a}                   & ---     & -49.650\\
$V_{l l}$ (MeV)       &  -1.01  &  -0.30\\
$V_{l s}$ (MeV)       &   7.00  &   7.00\\
$V_{ss}$ (MeV)        &  -0.20  &  -1.45\\
\end{supertabular}

\begin{supertabular}{>{\centering}p{12mm} p{13mm}<{\centering} 
>{\centering}p{13mm} p{10mm}<{\centering} p{13mm}<{\centering} }
\hline
\hline\\[-1.9ex]
$R_0$ & $a_0$ & $\beta_2$ & $\overline{\beta_2}$ & $\beta_4$\\
3.1 fm & 0.75 fm & 0.22 & 0.1034 & -0.08\\
\end{supertabular}
\begin{supertabular}{>{\centering}p{33mm} p{7mm}<{\centering} 
p{7mm}<{\centering} p{7mm}<{\centering} p{7mm}<{\centering}}
\hline
\hline\\[-1.9ex]
  &$1s_{1/2}$ &$1p_{3/2}$ &$1p_{1/2}$&$1d_{5/2}$\\
\hline\\[-1.9ex]
 $\lambda^{(OPP)}$ & 10$^6$ & 10$^6$ & 10$^6$ & 0.0\\
\hline
\hline\\[-1.9ex]
\end{supertabular}
$\overline{\beta_2}$ for linking $2^+_2$ and $4^+_2$ to each other and other
states; 43\% of 0.22.
\end{table}

There are three higher-spin states in the MCAS spectrum that have no
currently known partner: a $\frac{9}{2}^+$ state at -5.061~MeV
relative to the neutron emission threshold, a $\frac{13}{2}^-$ state
at -3.565~MeV, and an $\frac{11}{2}^-$ state at -1.764~MeV. If a
$\frac{9}{2}^+$ state existed at such low energy, it would likely have
been observed in experiment by population of $E1$ $\Gamma$-emission
from the $\sfrac{7}{2}^-_1$ state predicted in this work. This
$\sfrac{7}{2}^-_1$ state is likely to exist near this energy, given
the observation of such a state in $^{23}$Ne.  In this calculation,
the state in question arises from coupling of the $^{22}$Mg $2^+$ and
$4^+$ states with protons of $\ell \leq $ 11, as
appropriate. To determine what partial probabilities the various
coupled-channel components contribute requires calculation of
co-ordinate space wavefunctions, which is future work. As it is known
from Ref.~\cite{Ba11} that three of these states of $^{22}$Mg are
involved in the ground state of $^{23}$Al, it is reasonable to expect
that a low-energy $\frac{9}{2}^+$ state exists, but with a higher energy
than that calculated. This likely indicates the need for a
more-refined scattering potential.

While there is room for further elaboration of this potential,
including better accounting for the difference in inter- and
intra-band coupling strengths, and consideration of Pauli
hindrance~\cite{Am13}, this potential suffices for making predictions
of the cross section of the mirror system, $p+^{22}$Mg, and the
spectrum of the compound system, $^{23}$Al. For this, a suitable
charge distribution of $^{22}$Mg is necessary.

\section{The Coulomb interaction for the ${\rm p}$+${\rm ^{22}Mg}$ system}
\label{charge}

Experimental measurement of the root-mean-square (rms) charge radius
of the stable and long-lived nuclei was mostly completed in the 1980s,
and parameterisations of charge distribution functions for these are
available~\cite{Vr87}. These have recently been used in MCAS
studies~\cite{Fr15,Sv15}. However, such experimental guidance is not
yet available for radioactive ion beam (RIB) nuclei, though the SCRIT
experiment at the RIKEN RIB Factory~\cite{Su09,Su14}
has begun taking data.

\begin{figure}[htp]
\begin{center}
\scalebox{0.4}{\includegraphics*{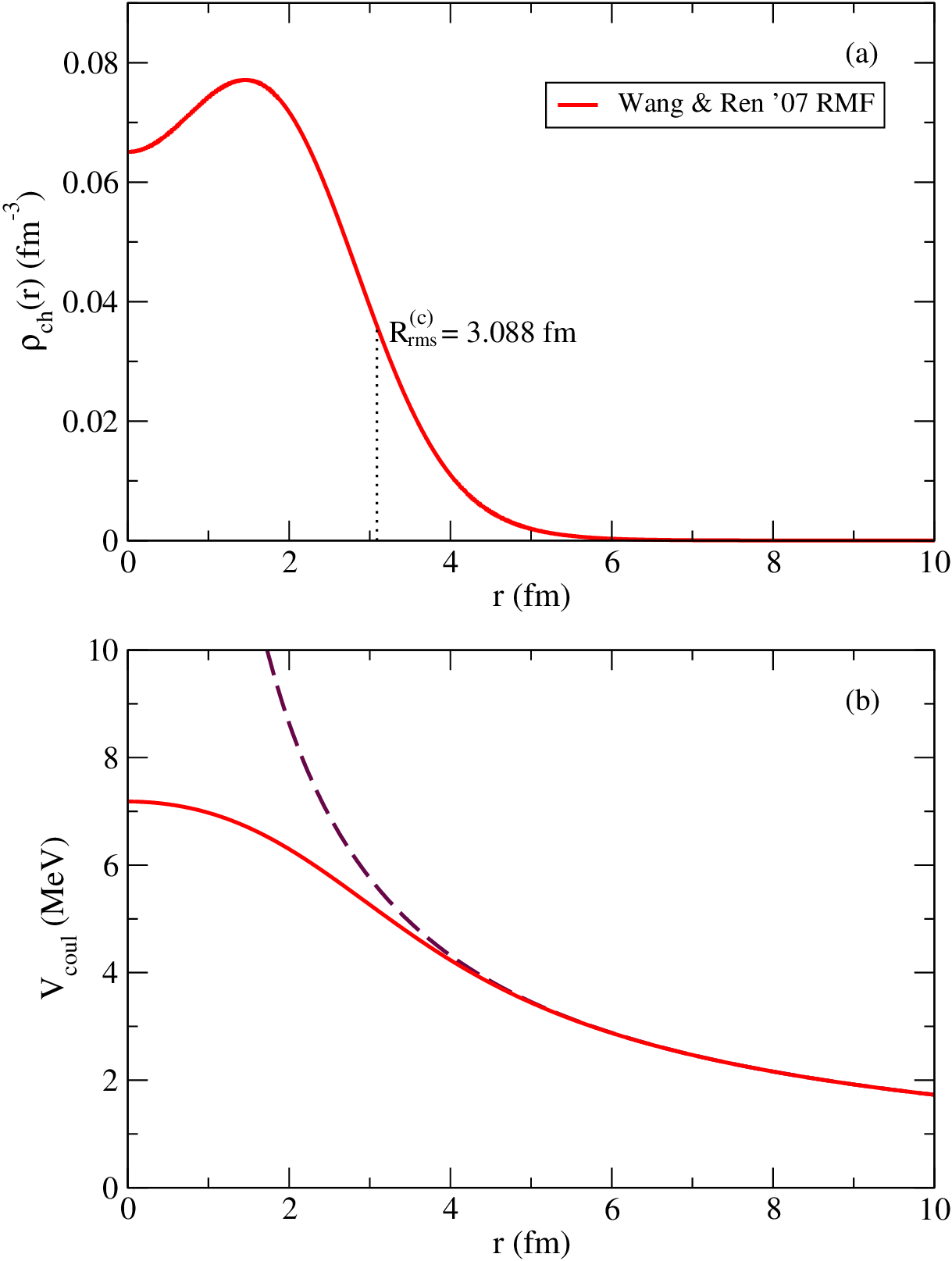}}
\end{center}
\caption{ \label{Wa07_dist} (a) RMF model theoretical charge
  distribution for $^{22}$Mg from Ref.~\cite{Wa07}. (b) Resultant
  Coulomb potential (solid line) and potential assuming point-like
  target (dashed line).}
\end{figure}

In the interim, charge-distribution information from other theories
may be used in generating Coulomb potentials for calculations of
proton-nucleus cluster structure and proton scattering with these
nuclei. One such theory, the relativistic mean-field (RMF)
model~\cite{Se86}, was used recently by Wang and Ren~\cite{Wa04,Wa07}
to calculate charge distributions for light proton-rich nuclei, the
properties of which they later compared with standard parameterised
functions of charge distribution~\cite{Ch10}. In particular, they
found an RMF charge distribution for $^{22}$Mg with an rms charge
radius, $R_{rms}^{(c)}$, of 3.088 fm, and this is used here to derive
the Coulomb potential for $p+^{22}$Mg scattering. This charge
distribution and the Coulomb potential are shown in panel (a) and (b)
of Fig.~\ref{Wa07_dist}, respectively. Additionally, panel (b) shows
the Coulomb potential that results from assuming a point-like
$^{22}$Mg target, i.e., $V_{Coul}~=~12~e^2~/~r$. This indicates
that the asymptotic behaviour of the potential derived from the charge
distribution is correct.

In~\ref{App-3pF}, the use of Coulomb potentials generated from
three-parameter Fermi (3pF) charge distributions is investigated, and
compared to the results from the RMF charge distribution.

\section{Observables of the ${\rm p}$+${ \rm^{22}Mg}$ system} 
\label{Al}

The compiled spectrum of $^{23}$Al consists of eight
states~\cite{Fi07}, of which only the ground state is bound, with the
one-proton separation energy being 123 keV. The ground state and first
known excited state at 0.55~MeV have angular momentum and positive
parity firmly assigned, as well as measured half-lives. Five more
states have firmly assigned positive parity and tentative angular
momentum. Only the excitation energy is known of the remaining
state. Since that compilation, He \textit{et al.}~\cite{He07} found
four resonances in $^1$H$(^{22}$Mg$,p)^{22}$Mg scattering with $E_x$ =
3.00, 3.14, 3.26 and 3.95~MeV. By $R$-matrix analysis, they argue that
the first and last of these may have spin-parity $\sfrac{3}{2}^+$ and
$\sfrac{7}{2}^+$, respectively. The state at 3.00~MeV is a likely
partner to the $\sfrac{3}{2}^+$ state in $^{23}$Ne at 3.432~MeV above
the ground state. The states at 3.14 and 3.26~MeV were
both tentatively assigned as $(\sfrac{7}{2}^+,\sfrac{5}{2}^+)$. Gade
\textit{et al.}~\cite{Ga08} measured a state 1.616~MeV above the
ground state which they argued to have spin-parity $\sfrac{7}{2}^+$,
of which a probable partner, currently assigned
$\left(\sfrac{5}{2},\sfrac{7}{2}^+\right)$, is known in $^{23}$Ne
1.701~MeV above the ground state.  Certainly, though, the low-energy
spectrum will be much richer than that so far observed.

\subsection{Charge symmetry}
\label{spec-Wa07}

\begin{figure}[htp]
\begin{center}
\scalebox{0.6}{\includegraphics*{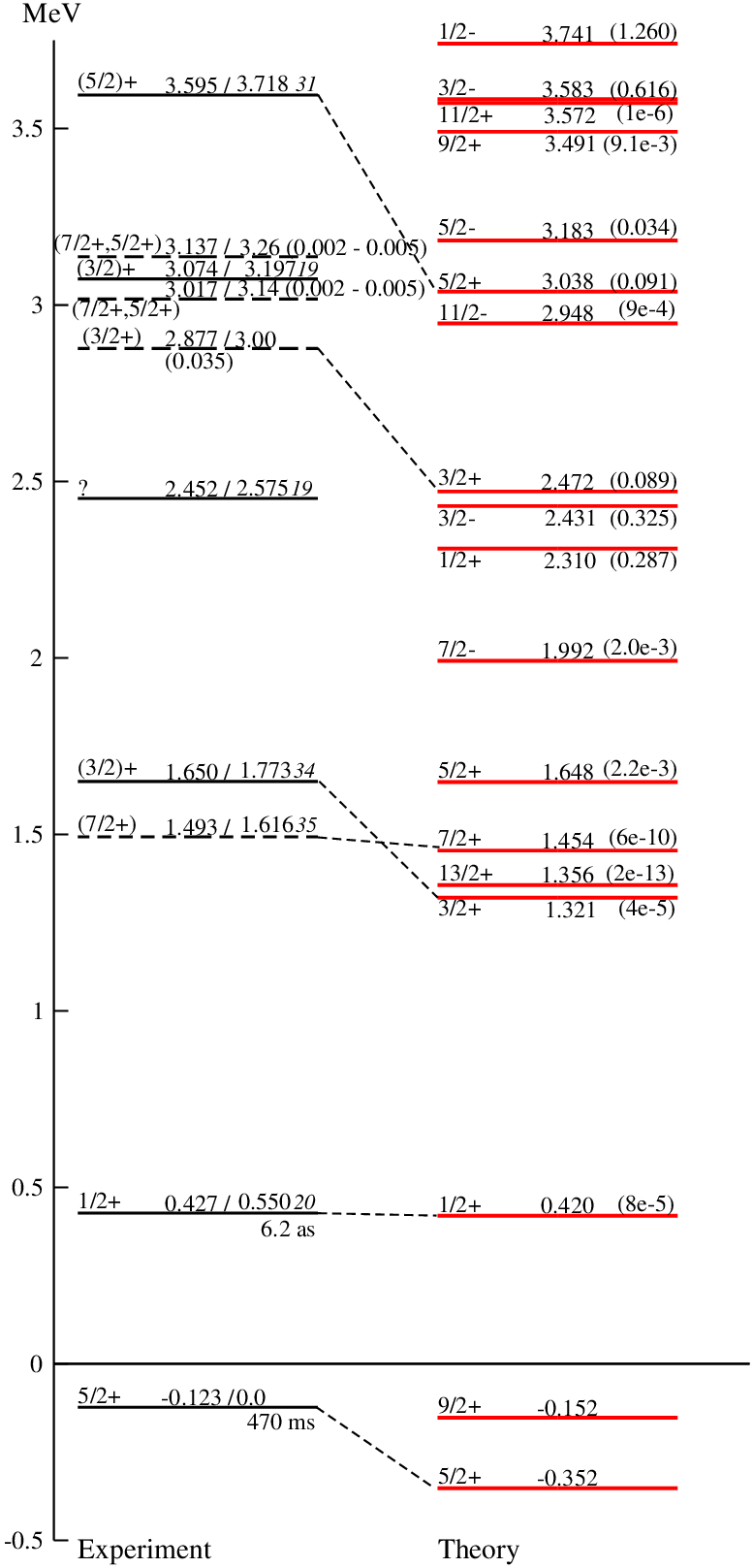}}
\end{center}
\caption{ \label{Al23-spec-Wa07} (Color online.) The known
  experimental $^{23}$Al spectrum (solid lines~\cite{Fi07}, long
  dashed lines ~\cite{He07}, short dashed lines~\cite{Ga08})
  and that calculated from MCAS evaluation of the $p+^{22}$Mg, with
  target state set $0^+_1$, $2^+_1$, $4^+_1$, $\overline{2^+_2}$ and
  $\overline{(4)^+_2}$, with charge distribution from
  Ref.~\cite{Wa07}.  Parameters are as per Table~\ref{params}, with
  $V_0^-~=~-65.200$~MeV and $V_0^+~=~-50.894$~MeV.  The bar denotes
  the use of reduced coupling for channels involving this state. In
  the experimental spectrum, energies before the slash are relative to
  the one-proton emission threshold, and energies after are relative
  to the ground state. Uncertainties are shown in italics. In the `Theory'
  spectrum, unbracketed values are level energy relative to one-proton
  emission threshold. Bracketed values are full width at half
  maximum. All energies are in MeV.}
\end{figure}

Fig.~\ref{Al23-spec-Wa07} shows the MCAS spectrum of $^{23}$Al that
results from using the parameter set of Table~\ref{params}, with the
addition of the Coulomb potential derived from the charge distribution
of Ref.~\cite{Wa07}. This is compared to the few observed states of
$^{23}$Al. Each experimentally observed state has a theoretical
partner nearby, though all are overbound. The ground state is
overbound by $\sim$229~keV. A deviation of similar magnitude has been
observed from studies of other mirror systems with MCAS calculations
made using Coulomb interactions defined from charge distributions with
the known rms charge radii~\cite{Fr15}. In those cases the deviation
could possibly be ascribed to nuclear effects (possibly charge
symmetry breaking). See \ref{App-3pF} for a study of the impact of the
choice of Coulomb potential used.

\subsection{Adjusted charge symmetry}
\label{spec-Wa07mod}

\begin{figure}[htp]
\begin{center}
\scalebox{0.6}{\includegraphics*{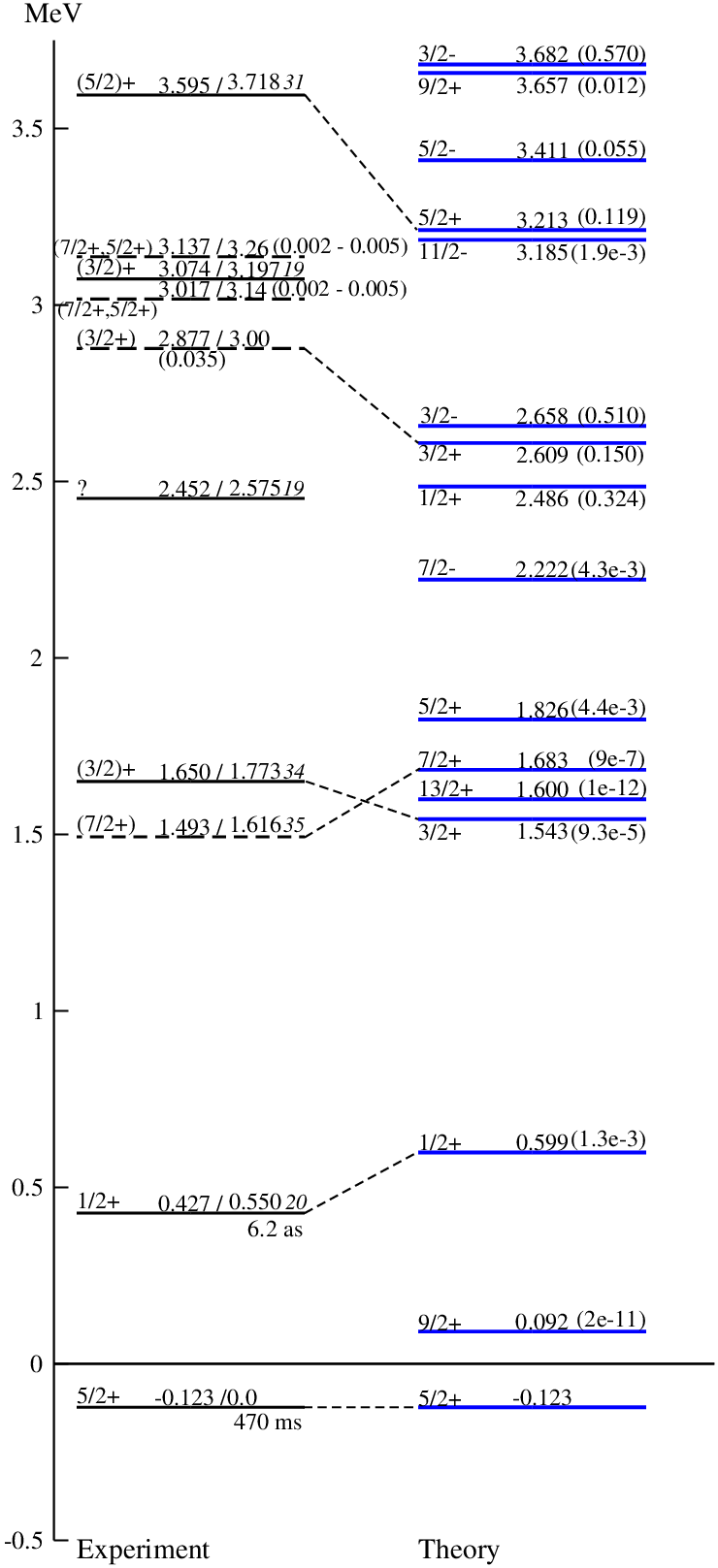}}
\end{center}
\caption{ \label{Al23-spec-Wa07mod} (Color online.) The known
  experimental $^{23}$Al spectrum~\cite{He07,Ga08,Fi07}, and that
  calculated from MCAS evaluation of the $p+^{22}$Mg, relative to the
  one-proton emission threshold. The MCAS results are obtained with
  target state set $0^+_1$, $2^+_1$, $4^+_1$, $\overline{2^+_2}$ and
  $\overline{(4)^+_2}$, with charge distribution from
  Ref.~\cite{Wa07}. Parameters are as per Table~\ref{params}, with
  $V_0^-~=~-64.620$~MeV and $V_0^+~=~-50.372$ MeV.  The bar denotes
  the use of reduced coupling for channels involving this state. In
  the experimental spectrum, energies before the slash are relative to
  the one-proton emission threshold, and energies after are relative
  to the ground state. Uncertainty in italics. In the `Theory'
  spectrum, unbracketed values are energy levels relative to
  one-proton emission threshold. Brackets values are full widths at
  half maximum. All energies are in MeV.}
\end{figure}

As the $p+^{22}$Mg central potential depth, $V_0$, is much greater
than the other components of the nuclear potential, slight adjustments
can be used to alter the binding of all calculated compound states
without altering the spacing between them. Using the RMF-derived
charge distribution from Ref.~\cite{Wa07}, a value of -50.372~MeV for
the central well depth for positive parity states, $V_0^+$, reduces
the ground state binding by the 229 keV necessary to obtain a match
with data. For the same increase in energy of the first calculated
negative parity state, with $J^\pi = \sfrac{7}{2}^-_1$, a value of
-64.620~MeV is used for the negative parity states' central well
depth, $V_0^-$. The resultant spectrum is shown in
Fig.~\ref{Al23-spec-Wa07mod}.

Thus, MCAS recreates the observed $\sfrac{5}{2}^+_1$,
$\sfrac{1}{2}^+_1$, $\sfrac{3}{2}^+_1$, $\sfrac{7}{2}^+_1$,
$\sfrac{3}{2}^+_2$ states, and the $\sfrac{5}{2}^+$ state observed at
3.595~MeV above the one-proton emission threshold, though the spins of
all but the lowest two are uncertain. The model gives many more
resonance states in the low excitation region and it remains to be
seen if such have any empirical matches. The resonance widths found
are solely for one-proton emission, which is not problematic as the
energies considered are below the emission thresholds for
$\alpha$-particles (8.58~MeV above the ground state), deuterons
(17.28~MeV), neutrons (19.48~MeV) and tritons (25.75
MeV). $\Gamma$-decay channels, while open, have negligible widths
compared to particle emissions. This is illustrated, for example, for
the $\sfrac{1}{2}^+_1$ state by Table IV of Ref.~\cite{Ca01}.

The model predicts a $\sfrac{5}{2}^+$ state (in
Fig.~\ref{Al23-spec-Wa07mod} at 1.826~MeV) not observed thus-far, in
addition to the ground state and putatively assigned
$\left(\sfrac{5}{2}\right)^+$ state observed at 3.595~MeV above the
one-proton emission threshold. It also predicts, relative to that
threshold, a $\sfrac{7}{2}^-_1$ at 2.222 MeV, a $\sfrac{1}{2}^+_1$ at
2.486~MeV, and a $\sfrac{3}{2}^-_1$ state at 2.658~MeV. However, if
the other known states in the region are a guide, it might be expected
that those states exist $\sim$0.5~MeV higher than MCAS
predicts. Furthermore, with this potential MCAS predicts several
states around 3.5~MeV (relative to the one-proton emission threshold),
and two high-spin states at lower energies.  These may correspond to
the partially defined states observed in this region.  To obtain a
better match to the three highest energy states observed
experimentally, being two $(\sfrac{7}{2})^+$ states at 3.95~MeV and
4.2~MeV, and a $(\sfrac{5}{2})^+$ at 11.8~MeV, all relative to the
ground state, requires improvement to the model specification of the
channel coupling interactions and, possibly taking more states of
${}^{22}$Mg into account.

In Table~\ref{results}, the MCAS results are compared with
experimental data and those of other calculations reported in the
literature~\cite{Ga08,Yu14,Fi07,Ti05}.  The shell-model calculation of
Ref.~\cite{Yu14} concerned itself with only the four lowest-energy
states known in $^{23}$Al, $\sfrac{5}{2}^+_1$, $\sfrac{1}{2}^+_1$,
$\sfrac{3}{2}^+_1$ and $\sfrac{7}{2}^+_1$, making no predictions. We
tabulate their calculation which uses the USD*($V^{pn}_{T=1}$)
interaction, as this provided them their best match to experimental
data. (Note that in their Table IV they appear to have the labels of
the $\sfrac{3}{2}^+_1$ and $\sfrac{7}{2}^+_1$ states reversed.)
Ref.~\cite{Ti05a} calculated properties of the $\sfrac{1}{2}^+_1$
resonance, obtaining an excitation energy of 0.405~MeV, and a proton
width of 2.01$\times10^{-2}$~MeV and 9.19$\times10^{-2}$~MeV from a
multichannel and a single-channel microscopic cluster model,
respectively.  Ref.~\cite{Ti11} reports on calculations focused solely
on the ground state and $\sfrac{1}{2}^+_1$ state of $^{23}$Ne and
ground state of $^{23}$Al, obtaining agreement with experiment to
three decimal places. Their interest, however, was with capture
reactions and so they used a simple model which couples a
valence nucleon to a rotor core~\cite{Nu96}, without some of the
features of the MCAS model used herein, and with only the $0^+$, $2^+$
and $4^+$ states of the core-nucleus' primary rotor band.  Based on
the results of Ref.~\cite{Fr14} and Section~\ref{no2_2} of this work,
it may be interesting for them to repeat their investigation including
also the $2^+_2$ core state. Ref.~\cite{He07} reported four new states
between 3 and 4~MeV above the ground state, and provided tentative
$J^\pi$ designations based on measurement and on $R$-matrix fits,
Table~\ref{results} quoting the those suggested by
measurement. Finally, in Ref.~\cite{Ga08} a theoretical estimate of
the width of the $\sfrac{7}{2}^+_1$ state they measured was made using
an USDB shell model calculation.

Widths for the three lowest-energy resonances known, both experimental
and theoretical, are extremely small in all investigations. Regarding
the $\sfrac{1}{2}^+_1$ state, the result of Ref.~\cite{Ti05a} is of
the same order as the experimental result, where the MCAS result is
larger.  With MCAS, the excitation energy of this state is
overestimated by 0.17~MeV.  The microscopic cluster model~\cite{Ti05a}
underestimated the energy of that state by about the same amount.
This state is important in any study of resonant capture and, as such
reactions are planned for future MCAS studies, more detailed analyses
leading to more accurate properties for this state will be needed.
However, as the goal in the present work is to predict the spectrum
over several MeV, rather than focusing solely on the Gamow window,
this discrepancy is not of prime concern here. The excitation energies
of the first four observed states reported in Ref.~\cite{Yu14} are
closer to experiment than we have found here, and closer to experiment
than the $\sfrac{1}{2}^+_1$ result in Ref.~\cite{Ti05a}.
Ref.~\cite{Yu14} presents no information for higher-energy states.
The $(\sfrac{3}{2}^+)$ measured at 3.00~MeV in Ref.~\cite{He07} is the
only state of that work for which the MCAS calculation finds a
match. The width of this state found by MCAS is five times that of the
measurement of Ref.~\cite{He07}. The other states of Ref.~\cite{He07}
are not recreated as these are in an energy regime
that is higher than where this MCAS calculations is
most accurate.

\begin{table*}[htp]\centering

\caption{\label{results} Resonance energies, relative to ground state,
  and decay widths of states in ${}^{23}$Al determined from
  experiment, MCAS, and calculations in the literature. $E_x$ is the
  excitation energy, $\Gamma_p$ is the one-proton emission width. All
  energies are in MeV. The OXBASH \cite{Yu14} result is their best match to
  data. The two $\Gamma_p$ values from \cite{Ti05a} are their single-
  and multi-channel results, respectively.}  \scriptsize
\begin{tabular}{c|cc|cc|cc|cc|cc|c}
\hline
\hline 
\multicolumn{1}{c}{} & \multicolumn{2}{c}{Compiled data}
& \multicolumn{1}{c}{Exp.} & \multicolumn{1}{c}{Shell model}
& \multicolumn{2}{c}{Exp.}
& \multicolumn{2}{c}{MCAS}
& \multicolumn{2}{c}{Microscopic cluster}
& OXBASH\\
\multicolumn{1}{c}{} & \multicolumn{2}{c}{\cite{Fi07}}
& \multicolumn{2}{c}{\cite{Ga08}}
& \multicolumn{2}{c}{\cite{He07}}
& \multicolumn{2}{c}{Fig.~\ref{Al23-spec-Wa07mod}}
& \multicolumn{2}{c}{model \cite{Ti05a}}
& \cite{Yu14}\\
\hline
$J^\pi$ & $E_x$ & $\Gamma_p$ & $E_x$ & $\Gamma_p$ & $E_x$ & $\Gamma_p$ & $E_x$
& $\Gamma_p$ & $E_x$ & $\Gamma_p$ & $E_x$\\
\hline
$\sfrac{5}{2}^+$   & 0.0   & N/A &&&&&
0.0 &  N/A &&& 0.0\\[1.1ex]
$\sfrac{1}{2}^+$   & 0.55  & 7.4$\times 10^{-5}$ &&&&&
0.722 & 1.30$\times 10^{-3}$  & 0.405 & 2.01$\times 10^{-5}$  & 0.57\\[1.1ex]
&&&&&&&&&& 9.19$\times 10^{-5}$ &\\[1.1ex]
$(\sfrac{7}{2}^+)$ &       &        & 1.616 & 1.1$\times10^{-9}$ &&&
1.806 & 8.70$\times 10^{-7}$  &&& 1.56\\[1.1ex]
$(\sfrac{3}{2})^+$ & 1.773 &&&&&&
1.666 & 9.32$\times 10^{-5}$  &&& 1.70\\[1.1ex]
?                  & 2.575 &        &&&&
&\multicolumn{2}{c|}{see below line}&&&\\[1.1ex]
$(\sfrac{3}{2}^+)$ &&&&& 3.00 & 0.032 & 2.732 & 0.1504 &&&\\[1.1ex]
$(\sfrac{7}{2}^+,\sfrac{5}{2}^+)$ &&&&& 3.14 & 2 - 5 keV &&&&&\\[1.1ex]
$(\sfrac{3}{2})^+$ & 3.197 &&&&&&&&&&\\[1.1ex]
$(\sfrac{7}{2}^+,\sfrac{5}{2}^+)$ &&&&& 3.26 & 2 - 5 keV &&&&&\\[1.1ex]
$(\sfrac{5}{2})^+$ & 3.718 &      &&&&& 3.336 & 0.1192  &&&\\[1.1ex]
$(\sfrac{7}{2}^+)$ &&&&& 3.95 & 0.02 &&&&&\\[1.1ex]
$(\sfrac{7}{2})^+$ & 4.200 &        &&&&&
- & - &&&\\[1.1ex]
$(\sfrac{5}{2})^+$ & 11.78 &        &&&&&
- & - &&&\\[1.1ex]
\hline
$\sfrac{9}{2}^+$  &&&&&&& 0.215 & 2.2$\times 10^{-11}$ &&&\\[1.1ex]
$\sfrac{13}{2}^+$ &&&&&&& 1.723 & 1.3$\times 10^{-12}$ &&&\\[1.1ex]
$\sfrac{5}{2}^+$  &&&&&&& 1.949 & 4.42$\times 10^{-3}$ &&&\\[1.1ex]
$\sfrac{7}{2}^-$  &&&&&&& 2.345 & 4.32$\times 10^{-3}$ &&&\\[1.1ex]
$\sfrac{1}{2}^+$  &&&&&&& 2.609 & 0.325 &&&\\[1.1ex]
$\sfrac{3}{2}^-$  &&&&&&& 2.781 & 0.511 &&&\\[1.1ex]
\hline
\hline
\end{tabular}
\end{table*}

\subsection{Elastic scattering cross sections}

\begin{figure}[t]
\begin{center}
\scalebox{0.46}{\includegraphics*{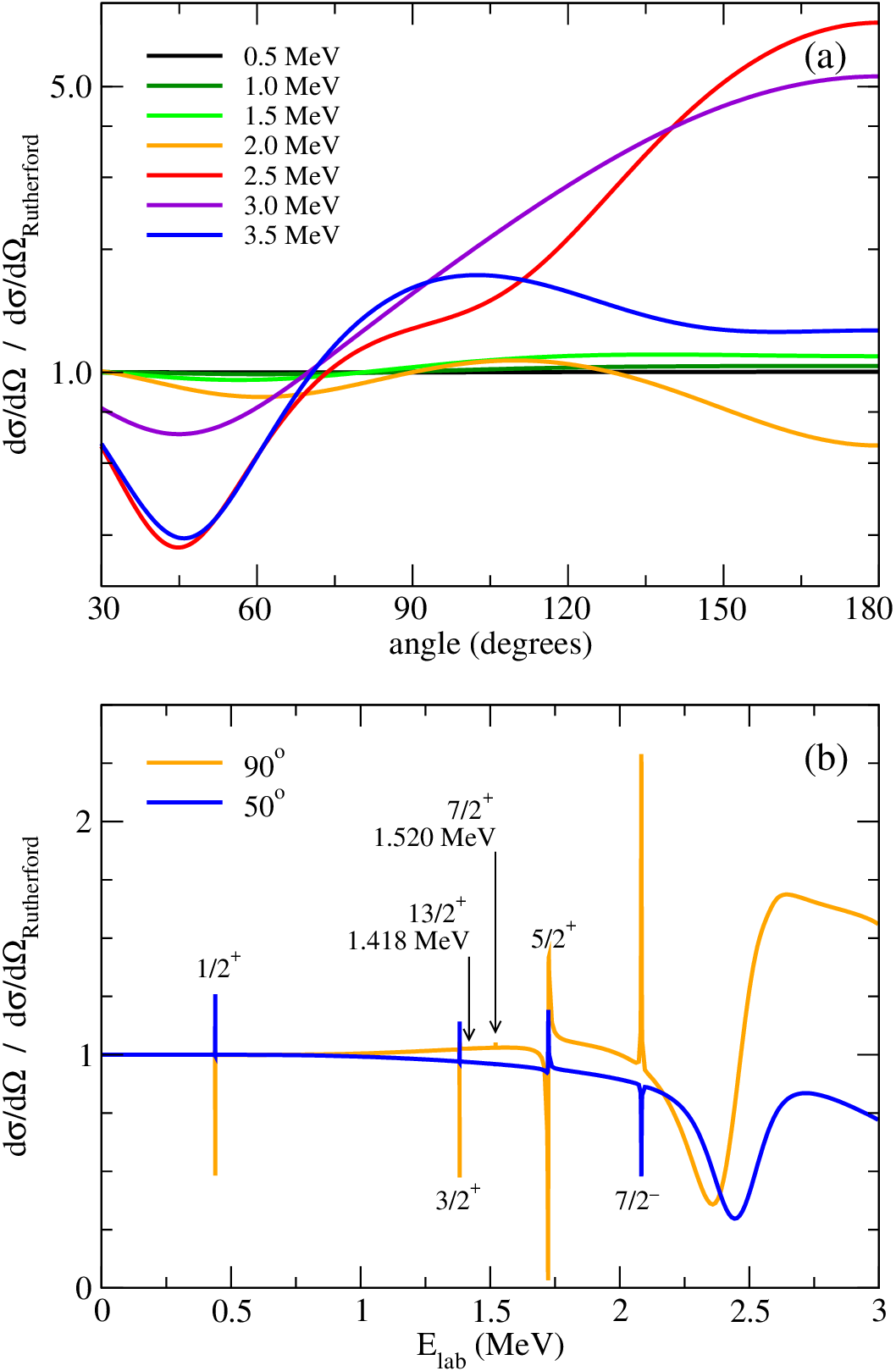}}
\end{center}
\caption{ \label{Wa07_xsec_rtR} (Color online.) (a) $p+^{22}$Mg
  elastic scattering cross section at several fixed proton laboratory
  energies (b) $p+^{22}$Mg elastic scattering cross section at fixed
  angles 50$^{\textrm{o}}$ and 90$^{\textrm{o}}$ for proton energies
  from 0.0 to 3.0~MeV. All presented as ratio to Rutherford cross
  sections.  Parameters are as per Table~\ref{params}, with
  $V_0^-~=~-65.200$~MeV and $V_0^+~=~-50.894$~MeV.}
\end{figure}

\begin{figure}[htp]
\begin{center}
\scalebox{0.46}{\includegraphics*{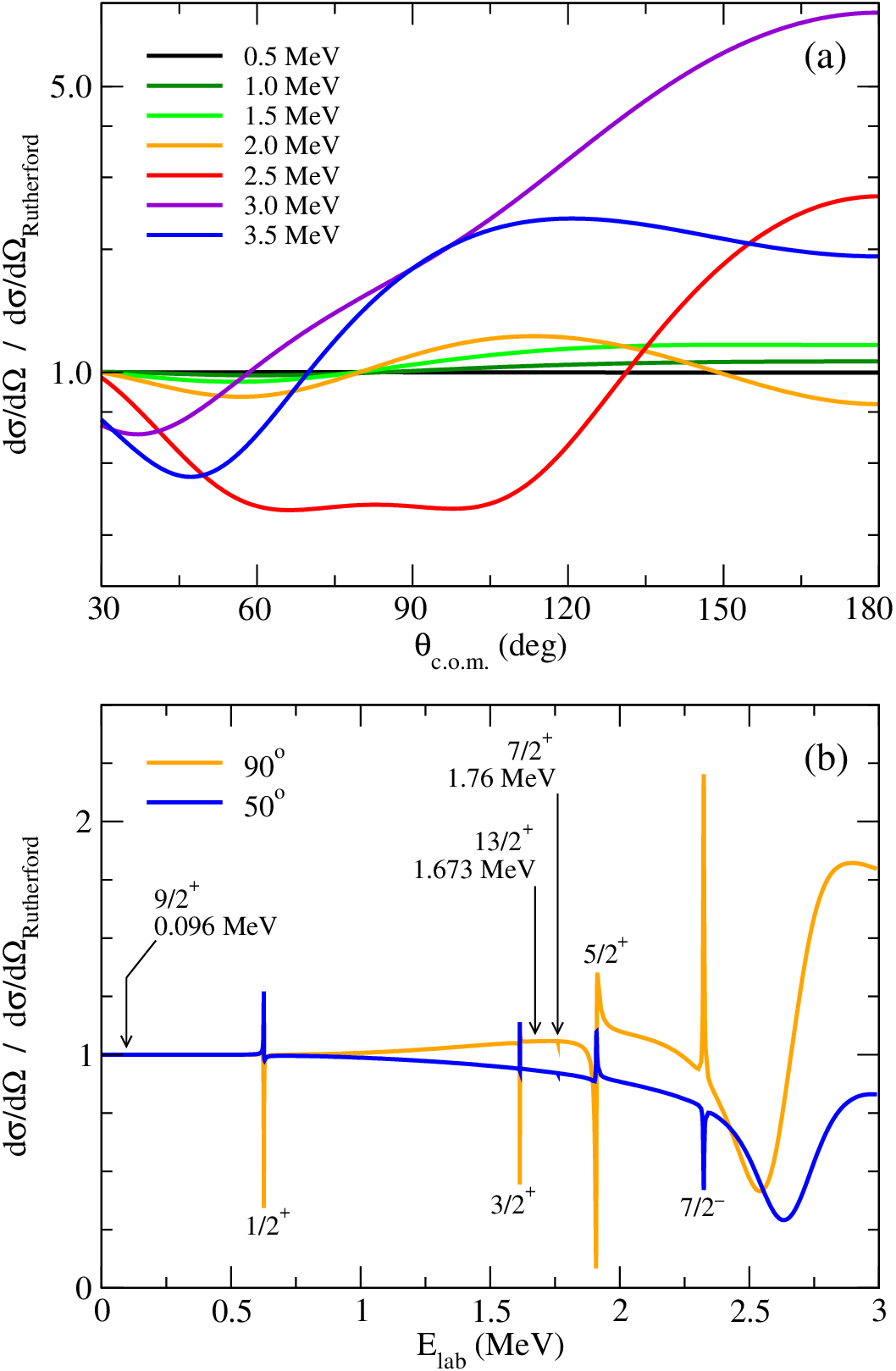}}
\end{center}
\caption{ \label{Wa07_xsec-mod_rtR} (Color online.) (a) $p+^{22}$Mg
  elastic scattering cross section at several fixed proton laboratory
  energies (b) $p+^{22}$Mg elastic scattering cross section at fixed
  angles 50$^{\textrm{o}}$ and 90$^{\textrm{o}}$ for proton from 0.0
  to 3.0~MeV. All presented as ratio to Rutherford cross sections.
  Parameters are as per Table~\ref{params}, with $V_0^-~=~-64.620$~MeV
  and $V_0^+~=~-50.372$~MeV.}
\end{figure}

Fig.~\ref{Wa07_xsec_rtR} shows several low-energy elastic scattering
cross sections obtained using the charge distribution of
Ref.~\cite{Wa07} and the same nuclear parameter set as for the
$n+^{22}$Ne investigation. Fig.~\ref{Wa07_xsec-mod_rtR} presents
cross sections that result from the adjustment of $V_0^+$ and $V_0^-$
detailed previously.

Comparing panels (a) in Figs.~\ref{Wa07_xsec_rtR} and
\ref{Wa07_xsec-mod_rtR}, the differential cross sections at 0.5, 1.0,
1.5 and 2.0~MeV, being in an energy region where only narrow compound
states are calculated, change little with the adjustment of state
excitation energies. The cross sections at 2.5, 3.0 and 3.5~MeV differ
dramatically, as would be expected from the number of compound states
with large widths calculated in this energy region and moved by the
change in parameters.

Comparing panel (b) in each figure, the cross sections remain very
similar until $\sim$2.25~MeV. Changing the $V_0^+$ parameter from
-50.894~MeV to -50.372~MeV removes the good agreement the
$\sfrac{1}{2}^+_1$ excitation energy had with experiment in
Fig.~\ref{Al23-spec-Wa07}, but in the cross sections the corresponding
resonance is of extremely narrow width. This means that the
non-resonant part of the cross section calculated in this energy regime
in Fig.~\ref{Wa07_xsec-mod_rtR} might be a reasonable
prediction. Importantly, the unobserved $\sfrac{9}{2}^+$ and
$\sfrac{13}{2}^+$ states found in the spectral calculation do not
appear in the calculated cross sections. Above $\sim$2.25~MeV, the
larger-width $\sfrac{5}{2}^+$, $\sfrac{5}{2}^-$, $\sfrac{9}{2}^+$ and
$\sfrac{3}{2}^-$ resonances dominate the cross section, and therefore
their excitation energies relative to states that may or may not exist
here influence the quality of the predicted cross sections 
at these energies. It remains to be seen what resonant states
at 2.5~MeV and higher exist before the accuracy of the cross sections
calculated in this energy region may be assessed.

\section{The effect of the {\rm $2^+_2$} target state}
\label{no2_2}

\begin{figure}[htp]
\begin{center}
\scalebox{0.6}{\includegraphics*{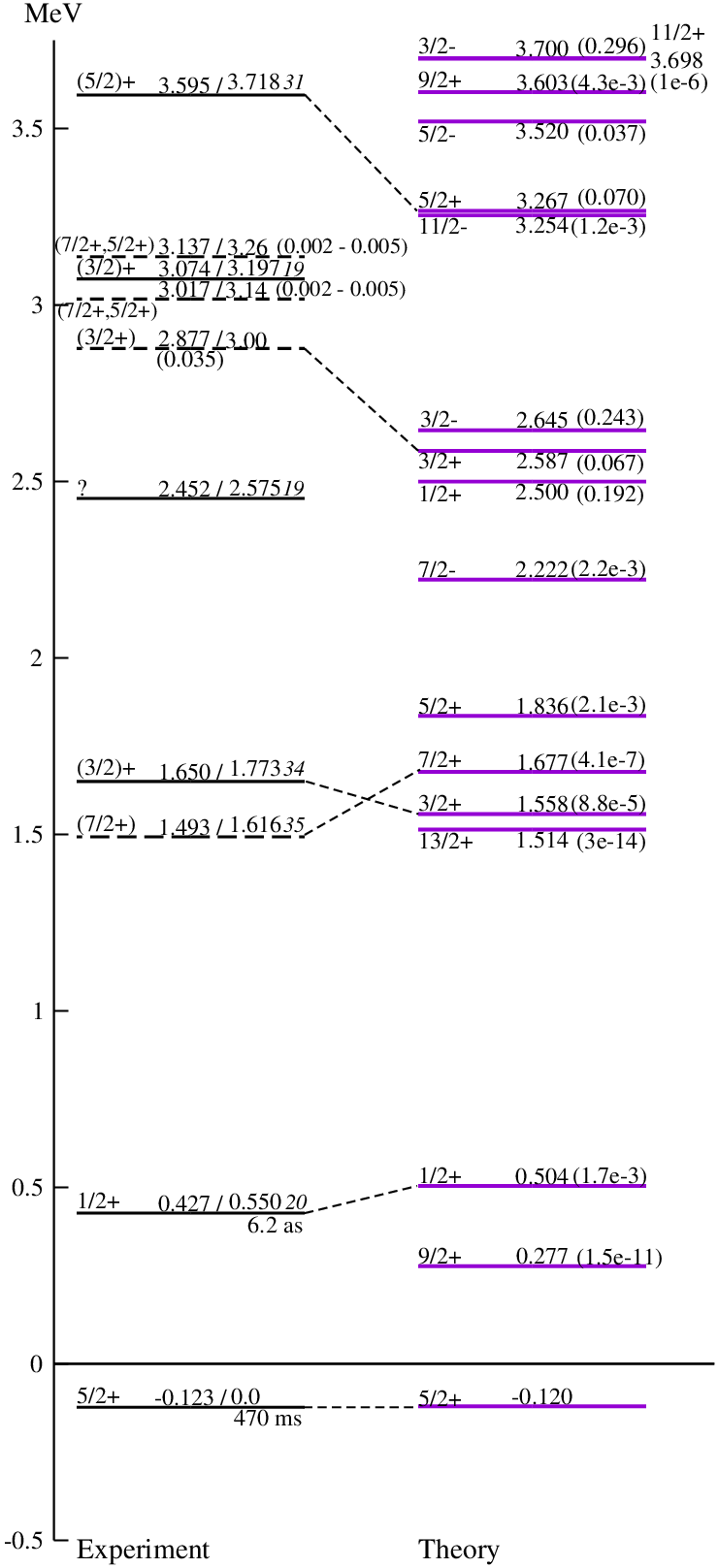}}
\end{center}
\caption{ \label{Al23-spec-Wa07no2_2} (Color online.) The known
  experimental $^{23}$Al spectrum~\cite{He07,Ga08,Fi07} and that
  calculated from MCAS evaluation of the $p+^{22}$Mg, relative to the
  one-proton emission threshold. The MCAS results are obtained with
  target state set $0^+_1$, $2^+_1$, $4^+_1$, and
  $\overline{(4)^+_2}$, with charge distribution from
  Ref.~\cite{Wa07}.  Parameters are as per Table~\ref{params}, with
  $V_0^-~=~-64.825$ and $V_0^+~=~-50.650$.  The bar denotes the use of
  reduced coupling for channels involving this state. In the
  experimental spectrum, energies before the slash are relative to the
  one-proton emission threshold, and energies after are relative to
  the ground state. Uncertainties are shown in italics. In the
  `Theory' spectrum, unbracketed values are energy levels relative to
  one-proton emission threshold. Bracketed values are full widths at
  half maximum. All energies are in MeV.}
\end{figure}

\begin{figure}[htp]
\begin{center}
\scalebox{0.46}{\includegraphics*{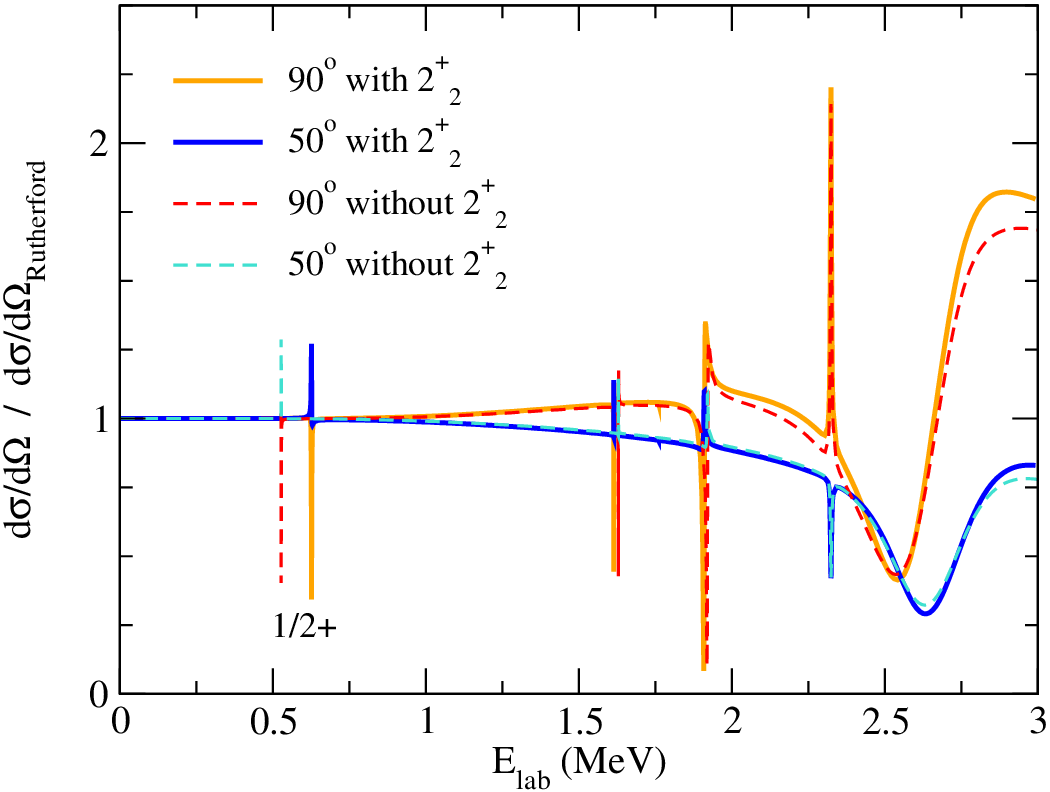}}
\end{center}
\caption{ \label{Wa07_xsec-no2_2_rtR} (Color online.) $p+^{22}$Mg
  elastic scattering cross section at fixed angles 50$^{\textrm{o}}$
  and 90$^{\textrm{o}}$ for proton energies from 0.0 to 3.0~MeV, presented as
  ratio to Rutherford cross sections.  Dashed lines are as per
  Table~\ref{params}, with $V_0^-~=~-64.825$~MeV and $V_0^+~=~-50.650$
  MeV, with target state set $0^+_1$, $2^+_1$, $4^+_1$, and
  $\overline{(4)^+_2}$. Solid lines are as per
  Fig.~\ref{Wa07_xsec-mod_rtR}.}
\end{figure}

When investigating which states of $^{22}$Mg couple with a valence
proton to form the $^{23}$Al ground state, Banu \textit{et
  al.}~\cite{Ba11} did not observe the $2^+_2$ state playing a
role. We repeat the calculation of Fig.~\ref{Al23-spec-Wa07mod} using
only the target-state set $0^+_1$, $2^+_1$, $4^+_1$, and
$\overline{(4)^+_2}$, to see if it erroneously plays a role in the
calculated ground state, and if it need be considered in other states
of $^{23}$Al. Again, $V_0^+$ was adjusted to match the ground state
to experiment. Without experimental guidance, $V_0^-$ was adjusted to
match the first predicted negative parity state, $\sfrac{7}{2}^-$, to
the same energy as in Fig.~\ref{Al23-spec-Wa07mod}. This gave the
values $V_0^-~=~-64.825$ MeV and $V_0^+~=~-50.650$~MeV. The resulting
spectrum is compared to experimental eigenenergies in
Fig.~\ref{Al23-spec-Wa07no2_2}. In Fig.~\ref{Wa07_xsec-no2_2_rtR},
the resulting fixed-angle cross sections are compared to those of
Fig.~\ref{Wa07_xsec-mod_rtR}.

It is seen that the $\overline{(2)^+_2}$ target state makes a
negligible contribution to the ground state, in agreement with
the experimental observation of Banu \textit{et al.}. 
It also makes a negligible contribution to all other
states in the calculated spectrum, with the following exceptions: its
absence brings the calculated counterpart of the $\sfrac{1}{2}^+$
resonance, observed at 0.427~MeV above the one-proton emission
threshold, into slightly better agreement with experiment; the energy
of the calculated $\sfrac{13}{2}^+$ resonance is slightly increased;
and the magnitude of cross sections is slightly reduced at the higher
end of the calculated energy range. Additionally, the calculated
widths of the resonances from this calculation are somewhat
smaller than those of Fig.~\ref{Al23-spec-Wa07mod}.

As the coupling of this state to the valence proton contributes to the
$\sfrac{1}{2}^+$ state just above the one-proton removal threshold,
its inclusion may be of importance for studies focused on that state,
for example Ref.~\cite{Nu96}.

\section{Conclusion}
\label{Conclusion}

Relatively little is known experimentally about the low-energy
spectrum of $^{23}$Al, and we have endeavoured to perform the most
comprehensive calculation of this spectrum to date, over several
MeV. We have treated it as a clusterisation of a proton and $^{22}$Mg,
using a coupled-channel approach. Channels have been defined by five
low excitation states in $^{22}$Mg.

The nuclear interaction was determined using mirror symmetry and the
related coupled-channel problem of the neutron+$^{22}$Ne
cluster~\cite{Fr14}.  The five states of the core nucleus were deemed
to be members of two rotation bands; the three lowest ($0^+_1$ g.s.,
$2^+_1$, $4^+_1$) belonging to a ground state rotation band. The
$2^+_2$ and $4^+_2$ were assumed as members of a second band with
coupling to the other three via a scaled strength. The deformation
strengths defining these couplings were linked to the known
$\gamma$-decay properties (in $^{22}$Ne).  The $2^+_2$ was found to
contribute only to the $\sfrac{1}{2}^+$ state of $^{23}$Al, supporting
the findings of Banu \textit{et al.} that it is not involved in the
ground state.  Coulomb interactions associated with a theoretical
model of the charge distribution of ${}^{22}$Mg~\cite{Wa07} were
used. The coupled-channel problems were solved using the MCAS
procedure in which the Pauli principle effects were
accommodated.

Where previous theoretical calculations have restricted themselves to
just the ground state~\cite{Ti11}, the ground state and first known
resonant state~\cite{Ti05a}, or the four lowest energy states
known~\cite{Yu14}, in this investigation, many $^{23}$Al eigenstates
in the first $\sim$3.5~MeV of the spectrum have been calculated, and
low-energy elastic scattering cross sections have been presented at
several energies and two fixed angles. After accounting for a small
overbinding which may be due to charge symmetry breaking of the
nuclear forces, a theoretical partner was found within $\sim$0.6
MeV for each state of the compound system with experimentally assigned
$J^\pi$. Below $\sim$2.25~MeV, where placement of excitation
energies is good, calculated proton-emission widths are very small, as
are the one known experimentally and the other theoretical results
available~\cite{Ga08,Ti05a}. Being very small, these do not have much
influence on the scattering background, giving some confidence that
the presented cross section calculations are reasonable
predictions. Such was the case in past uses of MCAS for nucleons on
$^{12}$C~\cite{Am13,Sv06} and $^{14}$O~\cite{Ca06a}.

Obtaining the known spectrum of ${}^{23}$Al through the threshold
region as well as elastic scattering cross sections (protons from
${}^{22}$Mg) is necessary for planned future evaluations of the
capture process using MCAS~\cite{Ca08}, so that both the scattering
and capture states can be defined by a single Hamiltonian. The capture
of protons by $^{22}$Mg has pertinence in specific nucleosynthesis
problems. We intend, as future work, to build MCAS to evaluate capture
cross sections and, for the case of proton capture by ${}^{22}$Mg, an
interaction then will be needed that gives better energy values for
the ground and first excited states in ${}^{23}$Al.

\section*{Acknowledgments}

The authors gratefully acknowledge Z.~Wang and Z.~Ren for providing
numerical results from their published nuclear charge distributions,
and J.~J.~He and collaborators for providing their published
$p+^{22}$Mg cross section data. This work was supported by the
Australian Research Council. ASK acknowledges a partial support from
the U.S. National Science Foundation under Award No. PHY-1415656. LC
acknowledges funds from the Dipartimento di Fisica e Astronomia
dell'Universit\`a di Padova. SK acknowledges support from the National
Research Foundation of South Africa.

\appendix
\section{3pF charge distributions}
\label{App-3pF}

It is instructive to investigate also a simple parameterised
functional form for the charge distribution of $^{22}$Mg. We examine
several cases of the three-parameter Fermi distribution~\cite{Ho71},
as this form recreates the central `dip' seen also in the charge
distribution of Ref.~\cite{Wa07}. It has the form
\begin{equation}
\rho_{ch}(r) = \rho_0 \frac{1 + \frac{wr^2}{R^2}}
                          {1 + \exp\left(\frac{r - R}{a} \right)} \;.
\label{3pF}
\end{equation}
The three parameters in this form for the charge distribution are,
$R$, a Woods-Saxon radius, $a$, the diffusivity of that function, and
$w$, a modifying form scale value.  Here $\rho_0$ is the central
charge density value with which the volume integral of this
distribution leads to the charge of the nucleus (12 for ${}^{22}$Mg).

\begin{figure*}[htp]
\begin{center}
\scalebox{0.7}{\includegraphics*{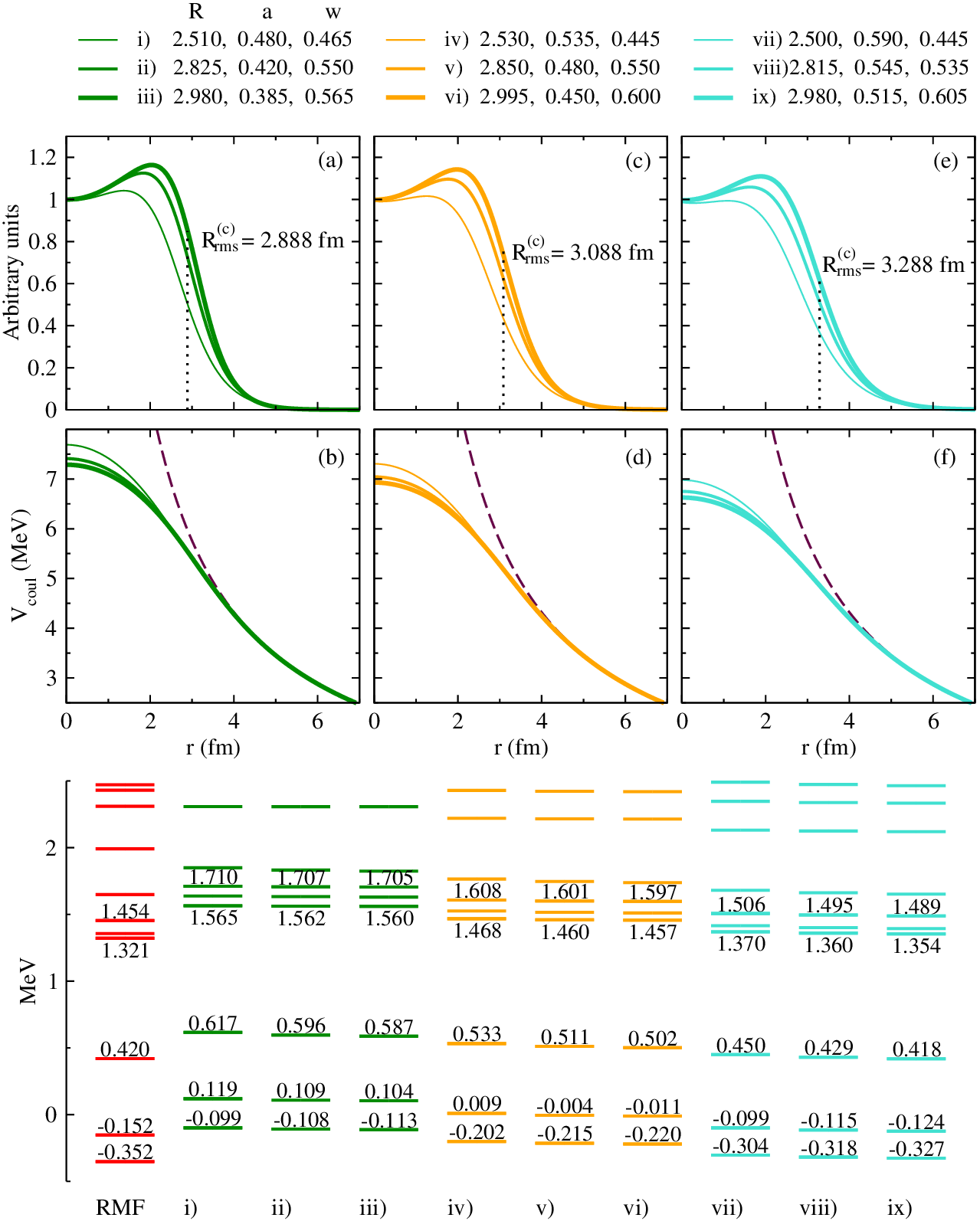}}
\end{center}
\caption{ \label{3pF_dist} (a) Three 3pF functions which result in
  charge distributions with $R_{rms}^{(c)} = 2.888$ fm.  (b) Resultant
  Coulomb potentials (solid lines) and potential assuming point-like
  target (dashed line).  (c) Three 3pF functions which result in
  charge distributions with $R_{rms}^{(c)} = 3.088$ fm.  (d) Resultant
  Coulomb potentials (solid lines) and potential assuming point-like
  target (dashed line).  (e) Three 3pF functions which result in
  charge distributions with $R_{rms}^{(c)} = 3.288$ fm.  (f) Resultant
  Coulomb potentials (solid lines) and potential assuming point-like
  target (dashed line). Bottom: Resulting MCAS spectra compared to
  that resulting from using the RMF charge distribution of
  Ref.~\cite{Wa07}.}
\end{figure*}

Normally, the parameters can be constrained to a surface in the
3-dimensional space by requiring the experimentally-known
$R_{rms}^{(c)}$. However, there is no experimentally
determined $R_{rms}^{(c)}$ value available for ${}^{22}$Mg. Panels (a), (c), and
(e) of Fig.~\ref{3pF_dist} each show three 3pF functions with
$R_{rms}^{(c)} = 2.888$ fm, $R_{rms}^{(c)} = 3.088$ fm (the value of
Ref.~\cite{Wa07}), and $R_{rms}^{(c)} = 3.288$ fm. We show the
functions in arbitrary units, with $\rho_0$ = 1 e$\cdot$fm$^{-3}$.
Panels (b), (d) and (f) show the Coulomb potentials that result from
the scaled functions, as well as the Coulomb potential assuming a
point-like target. The bottom panel of Fig.~\ref{3pF_dist} shows the
MCAS $^{23}$Al spectra that result from using these Coulomb potentials
with the nuclear potential used in Fig.~\ref{Ne23-spec} and
\ref{Al23-spec-Wa07}. The spectrum of Fig.~\ref{Al23-spec-Wa07},
obtained by using the RMF charge distribution of Ref.~\cite{Wa07}, is
also shown.

The different 3pF charge distributions for each value of
$R_{rms}^{(c)}$ result in similar spectra of $^{23}$Al.  As was
observed in Ref.~\cite{Fr15}, parameter variations essentially lead
only to an energy shift of the spectrum of a few tens of keV. Changes
in $R_{rms}^{(c)}$ result in greater changes in the spectra, of the
order of hundreds of keV for the values examined. This reaffirms that
when using a 3pF charge distribution, the $R_{rms}^{(c)}$ is the
important consideration. 
The exact values of the parameters ($R, a, w$) that coincide 
with the required $R_{rms}^{(c)}$ are of secondary importance.

Comparing the Coulomb potential in panel (b) of Fig.~\ref{Wa07_dist}
with those of panels (b), (d) and (e) of Fig.~\ref{3pF_dist}, we see
that the RMF-derived potential differs from those of a 3pF charge
distribution in two ways: it has a higher strength at $r = 0$ fm than
a potential from a 3pF distribution of lower $R_{rms}^{(c)}$, but has
lower strength at larger $r$, more like a potential derived from a 3pF
charge distribution of higher $R_{rms}^{(c)}$.

With these differences in potential we expect variations in calculated
observables, and indeed the RMF charge distribution results in lower
energies of states than the 3pF distributions with the same
$R_{rms}^{(c)}$, by $\sim$150 keV.  This of course means that using a
3pF distribution with this charge radius in Section~\ref{spec-Wa07mod}
would require less correction to the central well, $V_0$, of the
nuclear potential to find matches to experimentally-known $^{23}$Al
eigenenergies. However, rather than use the $R_{rms}^{(c)}$ value of
Ref.~\cite{Wa07} but not that charge distribution, we have opted to
use both in absence of experimental guidance.

\section{Observables of the ${\rm p}$+${ \rm^{22}Mg}$ system, 
         {\rm$E_{com}=$~2.6~MeV} to {\rm3.1~MeV}}
\label{App-Al2}

In Section~\ref{Al}, the parameters of the scattering potential were
selected to give the best possible fit to spectral data for the first
$\sim$2~MeV above the $^{23}$Al ground state, while predicting the
existence of eigenstates up to $\sim$3.5~MeV, if not their exact
energies.  Elastic cross sections were predicted for this region of
best fit, where no data currently exists.

Elastic scattering differential cross section data, however, do exist
for the energy window $E_{com}=$ 2.6 to 3.1~MeV, from
Ref.~\cite{He07}. In that study, recoiling protons from the reaction
$^1$H$(^{22}$Mg,$p)^{22}$Mg were detected at $\sim4^\circ$, $\sim17^\circ$,
and $\sim23^\circ$, which correspond in the centre of mass frame to
172$^\circ$, 142$^\circ$, and 134$^\circ$. It is for the first two of these, and
in the centre of mass frame, that elastic scattering cross sections
were provided.

\begin{figure}[htp]
\begin{center}
\scalebox{0.6}{\includegraphics*{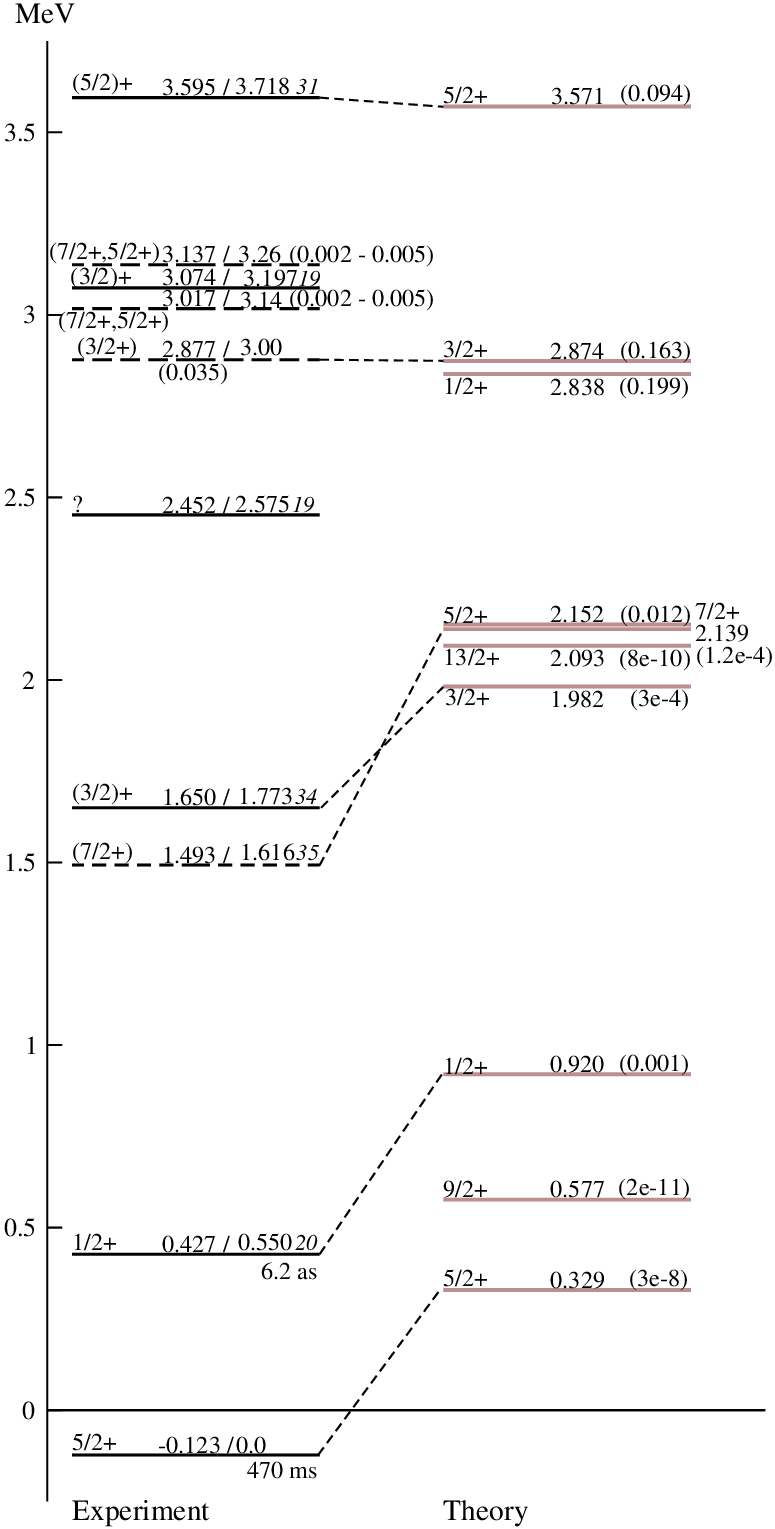}}
\end{center}
\caption{ \label{Al23-spec-He07a} (Color online.) The known
  experimental $^{23}$Al spectrum~\cite{He07,Ga08,Fi07} and that
  calculated from MCAS evaluation of the $p+^{22}$Mg, relative to the
  one proton emission threshold.  The MCAS results are obtained with
  target state set $0^+_1$, $2^+_1$, $4^+_1$, and
  $\overline{(4)^+_2}$, with charge distribution from
  Ref.~\cite{Wa07}.  Positive-parity are as per Table~\ref{params},
  with $V_0^+=-49.650$~MeV. No negative-parity potential is used. The
  bar denotes the use of reduced coupling for channels involving this
  state. In the experimental spectrum, energies before the slash are
  relative to the one-proton emission threshold, and energies after
  are relative to the ground state. Uncertainties are shown in
  italics. In the `Theory' spectrum, unbracketed values are energy
  levels relative to one-proton emission threshold.  Bracketed values
  are full widths at half maximum. All energies are in MeV.}
\end{figure}

\begin{figure}[htp]
\begin{center}
\scalebox{0.45}{\includegraphics*{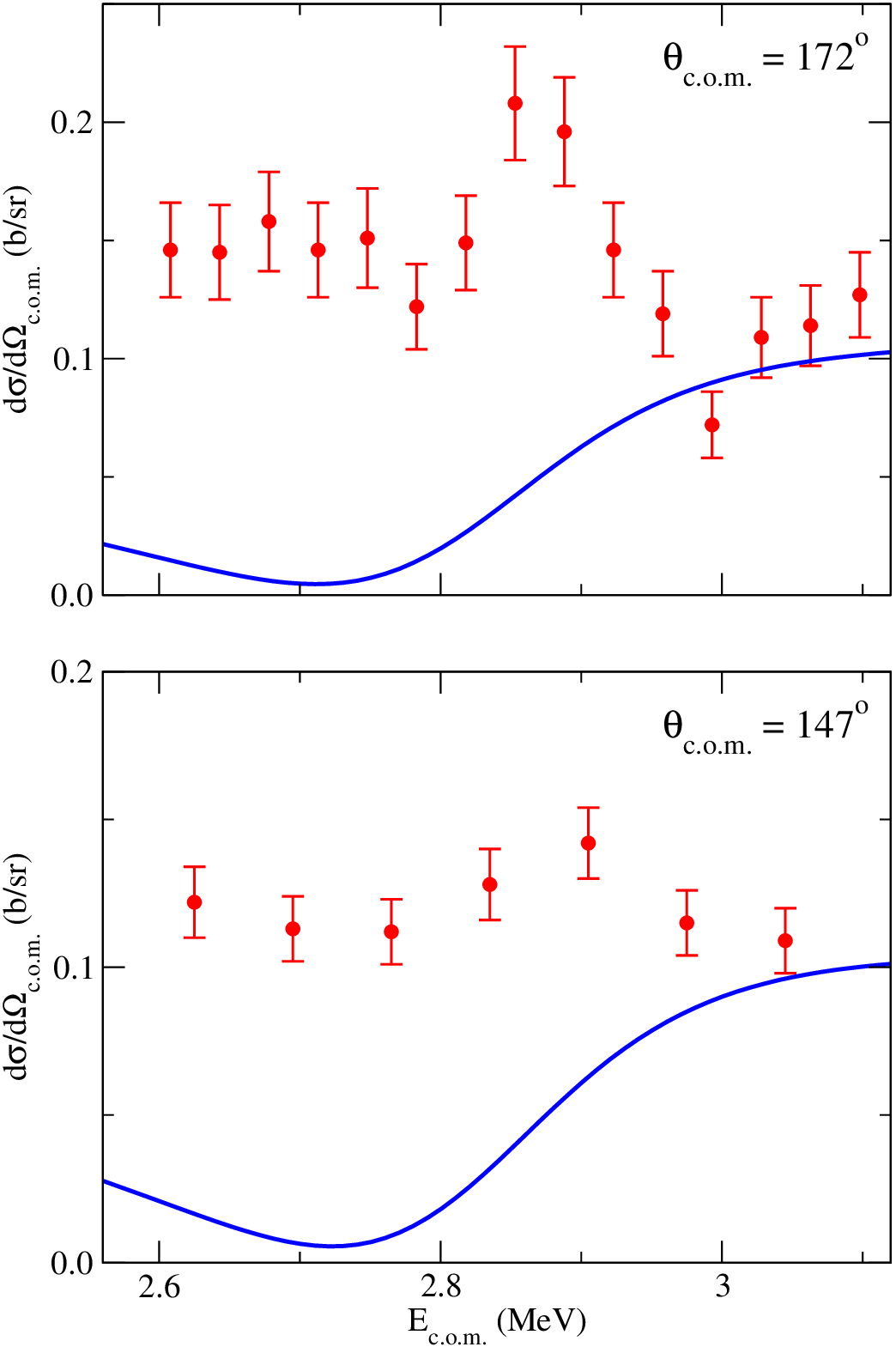}}
\end{center}
\caption{ \label{He07_xseca} (Color online.) $p+^{22}$Mg elastic
  scattering cross section for proton energies from 2.6 to 3.6~MeV at
  fixed angle (a) 172$^{\textrm{o}}$ and (b) 147$^{\textrm{o}}$ for
  proton energies from 2.6 to 3.1~MeV.  Parameters are as per
  Table~\ref{params}, with $V_0^+~=~-49.65$ MeV. The experimental data
  are from Ref.~\cite{He07}.}
\end{figure}

We stress that the potential used here is not best suited to this
energy window, where the density of states is higher and behaviour
other than that of a valence proton above a rotor-like core may affect
the spectrum. Additionally, there are almost certainly negative-parity
states in this region, but as none have been experimentally observed,
we cannot assess their placement by MCAS. However, we have adjusted
the $V_0^+$ well depth to match the MCAS $\sfrac{3}{2}^+_2$ state's
energy with that observed in Ref.~\cite{He07} at 2.877~MeV relative to
the one-proton emission threshold. Additionally, the negative-parity
potential was turned off to remove uncertainly from potentially
poorly-placed states. The resultant spectrum is shown in
Fig.~\ref{Al23-spec-He07a}, and the resultant cross sections are shown
in Fig.~\ref{He07_xseca}.

Given the conditions outlined above, the results are reasonable. The
resonance feature at $\sim$2.85~MeV in the data is of different shape than
that in the MCAS cross section, but the order of magnitude of the
cross section is the same, and the theoretical cross section passes
through or near several of the higher-energy data points.

\bibliography{Al23}

\end{document}